\documentclass[traditabstract,letter]{aa}  
\usepackage{graphicx}
\usepackage{txfonts}
\usepackage{bm}
\usepackage{natbib}

\bibpunct{(}{)}{;}{a}{}{,} 

\newcommand{\degree}{\ensuremath{^\circ}}
\newcommand{\phs}{\ensuremath{\phantom{-}}}

\begin{document}

\title{Discovery of a double eclipsing binary with periods near a 3:2 ratio}
\titlerunning{Discovery of a double eclipsing binary}
   \author{P. Caga\v{s}
          \inst{1}
          \and
          O. Pejcha
          \inst{2}
          }

   \institute{Modr\'a 587, 760 01 Zl\'in, Czech Republic, \email{pavel.cagas@gmail.com}
         \and
             Department of Astronomy, The Ohio State University, 140 W 18th Avenue, Columbus, OH 43210, USA    \\
		   \email{pejcha@astronomy.ohio-state.edu}
             }

  \date{Received ; accepted }

 
  \abstract
{The evolution of multiple stellar systems can be driven by Kozai cycles and tidal friction (KCTF), which shrink the orbit of the inner binary. There is an interesting possibility that two close binaries on a common long-period orbit experience mutually-induced KCTF. We present the discovery of a possible new quadruple system composed of two unresolved eclipsing binaries (EBs), CzeV343 ($V\!\sim\! 13.5$\,mag). We obtained photometric observations of CzeV343 that completely cover the two orbital periods and we successfully model the light curves as the sum of two detached EBs. We provide confidence intervals for the model parameters and minima timings by bootstrap resampling of our data. One of the EBs shows a distinctly eccentric orbit with a total eccentricity of about $0.18$. The two orbital periods, $1.20937$ and $0.80693$\,days, are within $0.1\%$ of a 3:2 ratio. We speculate that this might be the result of KCTF-driven evolution of a quadruple system and we discuss this hypothesis in the context of other quadruple systems composed of two EBs. We make our double EB fitting code publicly available to provide a tool for long-term monitoring of the mutual orbit in such systems.
}


   \maketitle


\section{Introduction}

The stability of multiple stellar systems requires that the stars are hierarchically organized \citep[e.g.][]{eggleton95,sterzik98} with an inner binary orbited by one or more outer bodies, although our knowledge of the formation and evolution of these systems is still limited. In the case of hierarchical triple systems, the presence of an outer body can drive the evolution of the inner binary through the Kozai cycles and tidal friction (KCTF) mechanism \citep[e.g.][]{mazeh79,kiseleva98,eggleton01,fabrycky07}. Kozai cycles periodically raise the eccentricity of the inner binary while the tidal friction efficiently dissipates the orbital energy during the close pericenter passages. The net result is a decrease of the orbital period of the inner binary. The KCTF can in principle explain the existence of close binaries with orbits smaller than their pre-main-sequence dimensions. Indeed, for orbital periods $P\lesssim 5$\,days the fraction of binary systems with a distant companion appears to increase \citep[e.g.][]{tokovinin06,pribulla06,dangelo06,rucinski07} and the eccentricity of the inner orbit is generally low \citep[e.g.][]{duquennoy91,raghavan10,dong12a}.

One prediction of the KCTF mechanism is the existence of a distant companion on typically high-inclination orbit around the inner binary. The parameters of the outer-body orbit can be obtained either directly via astrometry or spectroscopy \citep[e.g.][]{horn96,soderhjelm99}, by using the inner binary as a clock that orbits around a common barycenter \citep[e.g.][]{mayer90,borkovits96,gies12} or by combination of both \citep[e.g.][]{ribas02,zasche07}. Furthermore, the properties of the inner binary (such as the apparent inclination) can change due to the perturbations from the outer body, potentially leading even to a cessation of the eclipses \citep[e.g.][]{eggleton01,zasche12}. In all cases, an often heterogeneous set of data has to be analyzed, which requires correct estimation and appropriate treatment of observational uncertainties.

One of the possible arrangements of multiple systems are quadruple systems composed of two close binaries on a mutual long-period orbit. There is an interesting possibility that the two close binaries mutually influence their orbits through KCTF, although to our knowledge no theoretical studies of such systems have been performed. \citet{tokovinin03} suggested KCTF as an origin of high eccentricity in \object{41 Dra}, which forms a quadruple system with \object{40 Dra}. A potential wealth of information on these systems is available if both binaries exhibit eclipses. To date, only three such systems have been confirmed as SB4 binaries: \object{BV~Dra} and \object{BW~Dra} \citep{batten65,batten86}, \object{V994~Her} \citep{lee08} and \object{KIC~4247791} \citep{lehmann12}. The first two are visual binaries while the third is currently unresolved. Additionally, \citet{ofir08} proposed that \object{OGLE J051343.14$-$691837.1} is also an SB4 binary where the two binary periods are in an exact 3:2 ratio, but later investigations revealed only an SB2 with unexplained spectral changes related to the second period \citep{kolaczkowski10,rivinius11}. \object{OGLE-LMC-ECL-16549} is composed of two unresolved binaries with very different periods \citep{graczyk11}. There are two other quadruple systems with only one of the binaries favorably oriented to show eclipses \citep{shkolnik08,harmanec07}.

In this paper, we present the discovery of CzeV343, a double eclipsing binary similar to \object{V994~Her} and \object{KIC~4247791}, but with the periods of the two eclipsing binaries very close to a 3:2 ratio. We describe CzeV343 and our photometric data in Section~\ref{sec:observations}. In Section~\ref{sec:model}, we present a model of the observed light curves and in Section~\ref{sec:blend} we discuss the physical connection between the two binaries. In Section~\ref{sec:disc}, we discuss and summarize our results.

\section{Observations}
\label{sec:observations}

The photometric variability of a $V\!\sim\!13.5$\,mag star CzeV343\footnote{\object{GSC~02405-01886}; $\alpha$=$5^{\rm h}48^{\rm m}24\fs008$, $\delta$=+30\degree 57\arcmin 03\farcs 64 (J2000)} was detected during a search for new variable stars with $0.25$\,m $f/5.4$ Newtonian telescope equipped with G4-16000 CCD camera\footnote{Parameters of our detectors are given at {http://www.gxccd.com}.} and a coma-corrector located at a private observatory in the Czech Republic. Initially, no photometric filter was used to maximize the throughput because the camera is sensitive between $350$ and $1000$\,nm with a maximum at $550$\,nm, but later we obtained $V$ and $I_{\rm C}$ band photometry with a higher quantum efficiency G2-3200 CCD camera. Most of the exposures were $180$\,s. All images were calibrated with appropriate dark frames and flat fields created as a median of five individual dark and flat exposures. We performed differential aperture photometry using C-Munipack\footnote{{http://c-munipack.sourceforge.net/}}, which is based on DAOPHOT \citep{stetson87}. A nearby star of similar brightness and similar color (\object{GSC~02405-01305}) was chosen as a comparison star to minimize the effects of differential extinction. In total, we obtained $833$ useful photometric measurements in $14$ nights spanning $\sim 80$\,days before the star became unobservable as it moved into conjuction with the Sun. Our data are summarized in Table~\ref{tab:nights}.

\begin{figure}
\centering
\includegraphics[width=0.45\textwidth,clip]{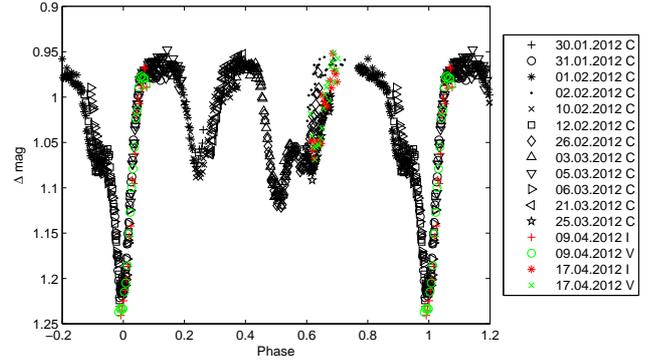}
\caption{Light curve of CzeV343 folded with the orbital period of system A of about $1.209$\,days. Each dataset is plotted with a different symbol explained in the right part of the plot.}
\label{phaseA}
\end{figure}

CzeV343 is a previously unknown variable star. Initially, CzeV343 appeared as a typical detached eclipsing binary in our data. However, subsequent observations revealed that the light curve is highly peculiar with three types of minima, as can be seen in the phased $\sim\!1.2$\,day period light curve shown in Figure~\ref{phaseA}. During the course of our observations, the minima at phases $0.25$, $0.6$ and $0.9$ were slowly drifting with respect to the primary and secondary minima. The only explanation that fits the observed periods and the light curve is that CzeV343 is composed of two eclipsing binaries with orbital periods of $1.2$ and $0.8$\,days, similar to \object{V994~Her} and \object{KIC~4247791}.


\section{Model of the light curves}
\label{sec:model}

In this section, we discuss the modeling of our data. The observed data consist of magnitudes $m_i$ and their uncertainties $\sigma_i$ measured at times $t_i$ distributed in $M$ datasets. Each dataset consists of $N_j$ measurements from one night and in one passband. We converted $t_i$ to the barycentric Julian dates in barycentric dynamical time (BJD$_{\rm TDB}$) using the on-line tool\footnote{{http://astroutils.astronomy.ohio-state.edu/time/utc2bjd.html}} of \citet{eastman10}. Similarly as \citet{lehmann12}, we modeled the observations as a sum of the fluxes of two detached eclipsing binaries,
\begin{equation}
m(t_i) = -2.5\log_{10} \left[F_A(\bm{\alpha}_A, t_i) + \beta F_B(\bm{\alpha}_B, t_i) + F_C \right] + \sum_{j=1}^M c_j \Delta_{ji},
\label{eq:model}
\end{equation}
where $F_A$ and $F_B$ are the fluxes from eclipsing binaries A and B scaled by the factor $\beta$, and $F_C$ represents any additional flux in the system. Each eclipsing binary is described by a set of parameters $\bm{\alpha}$. For each binary we varied the orbital period $P$, time of primary minimum $T_0$, the sum of radii $r_1+r_2$ measured relative to the semi-major axis, the ratio of radii $r_2/r_1$, the surface brightness ratio $\sigma$, the inclination $i$ of each orbit with respect to the observer, and for the binary with longer period we also vary two eccentricity parameters ($e \sin \omega$, $e \cos \omega$). We fixed the mass ratios of both systems to unity, because this has little effect for detached systems. Lacking the spectral types of the components and given the quality of the photometry, we set the gravity and linear limb darkening coefficients for all components and all filters to $0.5$. We allowed for magnitude shifts $c_j$ between the individual datasets and the model; $\Delta_{ji}$ is unity if the time instant $t_i$ is within the dataset $j$, and zero otherwise.

\begin{figure}
\centering
\includegraphics[width=0.45\textwidth]{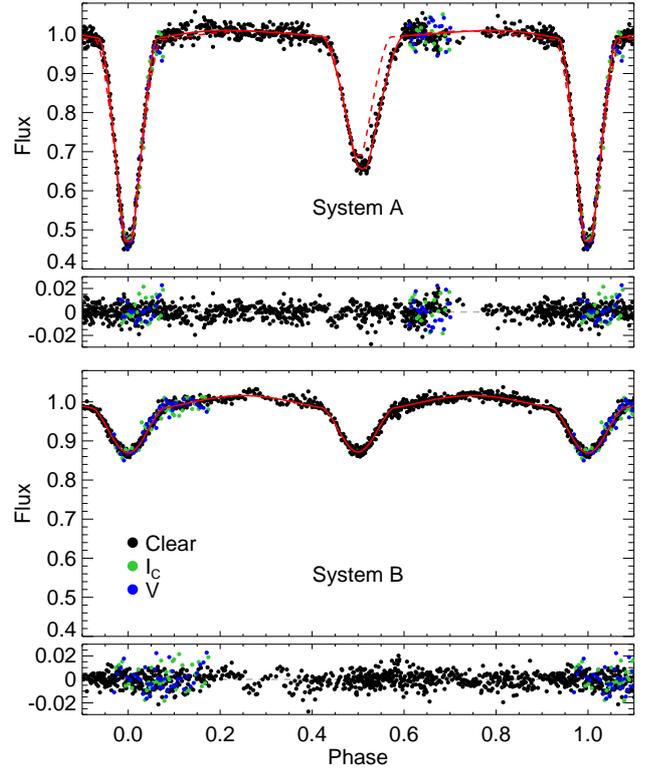}
\caption{Light curves of the two eclipsing binaries with the model contribution of the other component removed phased with the appropriate $P$ and $T_0$. The best-fit model (Table~\ref{tab:params}) is shown with solid red lines. The dashed red line shows best-fitting model with $e_A \equiv 0$. The smaller panels show the magnitude residuals phased with respect to the period of each of the systems.}
\label{fig:see_phase}
\end{figure}

We fit all parameters simultaneously by minimizing $\chi^2$ using the routine {\tt cmpfit}\footnote{{http://cow.physics.wisc.edu/$\sim$craigm/idl/cmpfit.html}} \citep{more78,markwardt09} with $F_A(\bm{\alpha}_A,t_i)$ and $F_B(\bm{\alpha}_B,t_i)$ computed with the code {\tt JKTEBOP} \citep{popper81,southworth04,southworth07,bruntt06}. We performed a number of bootstrap resamplings of the original data to derive reliable confidence intervals of the parameters. For both the original and bootstrapped data we started the minimization at a number of random positions in the parameter space to find the lowest $\chi^2$. 

We found that our model has degeneracies in several parameters. First, the scaling factor $\beta$ is to a large extent degenerate with the additional flux $F_C$ and several of the binary parameters such as inclination. We performed the minimization for a range of fixed values of $\beta$ with either $F_C$ free to vary or fixed at $F_C\equiv 0$. With $F_C$ left to vary, we found quite small $\Delta\chi^2 \approx 5$ for $0.6 \lesssim \beta \lesssim 1.8$. Because there is little prior information on $\beta$, we decided to fix $F_C\equiv 0$, which yields $\beta \approx 1.85$ with a second minimum at $\beta \approx 1.50$ differing by $\Delta\chi^2 \approx 10$. The assumption of equal fluxes $\beta = 1$ is $\Delta\chi^2 \approx 40$ worse than $\beta \approx 1.85$. In our fiducial solution, we set $F_C\equiv 0$ and $\beta\equiv 1.85$. There is also a degeneracy in $r_2/r_1$ in the sense that bootstrapping results occasionally yield two peaks: one with $r_2/r_1 < 1$ and the other with $r_2/r_1 > 1$. The relative probability of the two peaks depends on $\beta$. This degeneracy occurrs, because $r_2/r_1$ essentially sets the flatness of the primary and secondary minima, which are constrained only by a few datapoints, and on the amount of blended light of the other binary, which is controlled by $\beta$. For $\beta \equiv 1.85$, about $86\%$ of the bootstrap resamplings have $r_2/r_1 > 1$, which we then used to obtain confidence intervals on the parameters. Our best-fit model has $\chi^2 = 1040$ for $833$ measurements in $16$ datasets and $30$ free parameters. The model parameters are given in Table~\ref{tab:params} and the light curves of the two eclipsing binaries are shown in Figure~\ref{fig:see_phase}. The model is a good fit of the data.

\section{Physical connection between the binaries}
\label{sec:blend}

\begin{figure}
\centering
\includegraphics[width=0.45\textwidth]{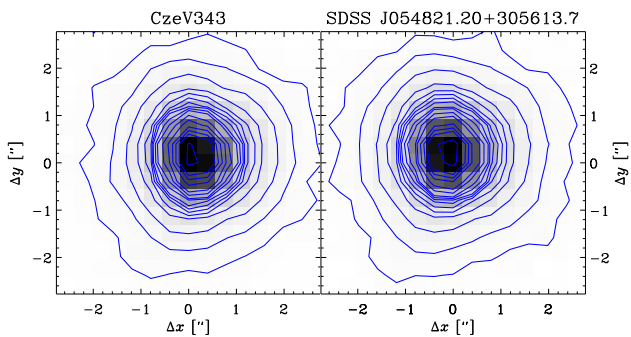}
\caption{Comparison of SDSS DR7 images of CzeV343 (left panel) and a similarly bright star about $1\arcmin$ away (right panel) with isophotal contours overplotted in blue. We show images in the $z'$ band, where CzeV343 does not saturate and the brightness difference between the two binary systems should be the smallest.}
\label{sdss}
\end{figure}

Even though a combination of two eclipsing binaries fits the data well, this does not prove that there is a physical connection -- the two binaries can simply be projected on the same position on the sky. To see whether CzeV343 can be resolved with our data, we analyzed centroid positions relative to a nearby star as a function of the flux. This method is used to constrain false positives in the search for transiting exoplanets \citep[e.g.][]{batalha10,jenkins10}. However, the plate scale of our setup is $1\farcs 39$/pixel with a typical stellar FWHM of $3$ pixels. Furthermore, our setup was designed for a wide field of view rather than for precise astrometry. As a result, we can only limit any shifts to be smaller than about $0\farcs 5$. Additionally, we analyzed the SDSS DR7 images \citep[plate scale of $0.396\arcsec$/pixel and a typical stellar FWHM of $\sim 2$\,pixels]{abazajian09} that have a resolution superior to our data. In Figure~\ref{sdss}, we show CzeV343 in comparison with a close star of similar brightness. There is no discernible difference between the two stars or any other stars nearby. More specifically, the measured PSF ellipticities and position angles are the same. CzeV343 is thus unresolved given the available data.


\section{Discussions and conclusions}
\label{sec:disc}

We showed that the light curve of CzeV343 is well described by a sum of the fluxes of two eclipsing binaries, but it is unclear whether the two eclipsing binaries are physically related. In the case of \object{KIC~4247791}, \citet{lehmann12} argued for a physical connection based on the similarity of masses, spectral types and periods of the four stars. Similar reasoning can be applied to CzeV343 as well, because the temperatures of the four stars must be similar given the lack of systematic differences between the $V$ and $I_{\rm C}$ band residuals (Fig.~\ref{fig:see_phase}). However, there is an observational bias against discovering blended double eclipsing binaries with significantly different temperatures without any regard to the  physical connection between the components. 

The question of whether systems like CzeV343 and \object{KIC~4247791} are truly quadruple systems can be established by obtaining accurate distances to all four stars, which requires precise photometry, radial velocities and spectral classifications. Alternatively, the physical connection can be proved by observing the effects of the gravitational interaction between the two systems with the added advantage of determining the parameters of the mutual orbit. In particular, light travel time effects in both systems with the same period would determine the mutual orbit. This method requires precise minimum timings with realistic uncertainties and with proper separation of the flux of both components. For example, there are over $50$ minima timings of V994~Her over the last $\sim 20$ years\footnote{{http://var.astro.cz/ocgate}}, but their practical usefulness is very low because in some cases simultaneously occurring minima were not properly separated. To start the long-term monitoring of CzeV343, we determined $O-C$ values for our datasets by fixing the binary parameters to the values in Table~\ref{tab:params} and fitting for magnitude shifts $c_j$ and changes in $T_0$. We bootstrap-resampled data in each dataset and repeated the fitting to give reliable $90\%$ confidence intervals. The results are presented in Table~\ref{tab:nights}. Because fitting such unusual light curves is not commonly done, we make our code publicly available\footnote{{http://www.astronomy.ohio-state.edu/$\sim$pejcha/czev343}} with the hope that this will allow determination of minima timings of CzeV343 and similar stars in the future. 

The physical parameters of the two binaries (Table~\ref{tab:params}) are not extraordinary in any way and are to some extent degenerate given our data, but it is worth pointing out that system A has an eccentric orbit. The secondary minima occur at phase $\sim 0.52$, but they also last longer than the primary minima, which yields a total eccentricity of $e_A \approx 0.18$. The best solution with $e_A \equiv 0$ is worse by $\Delta\chi^2 \approx 1360$ and is shown in Figure~\ref{fig:see_phase} with a dashed red line. An eclipsing binary with $e \approx 0.18$ and $P\sim 1.2$\,days is quite rare \citep{devor05}. Observations of the apsidal motion in system A and orbital precession can constrain the parameters of the mutual orbit as well \citep{eggleton01}. 

Finally, the most interesting property of CzeV343 is that the two orbital periods are very close to a 3:2 ratio:
\begin{equation}
1-\frac{2P_A}{3 P_B} = (8.5 \pm 0.2) \times 10^{-4}.
\label{eq:resonance}
\end{equation}
The difference is visible in Figure~\ref{phaseA} as a slow drift of system B minima. Given their relatively small orbital separations, systems such as CzeV343 likely experienced KCTF that shrunk their orbits. There is a possibility that during this process the orbital periods became locked in a 3:2 resonance. This interpretation is appealing because the light curve of \object{OGLE J051343.14$-$691837.1} also appears as a combination of two contact binaries with periods in an exact 3:2 ratio, although this has not been confirmed yet through spectroscopy \citep{ofir08,kolaczkowski10,rivinius11}. The periods in \object{V994~Her}, $2.08$ and $1.42$\,days, are considerably farther from 3:2 with a calculation similar to Equation~(\ref{eq:resonance}) giving $\sim 2\times 10^{-2}$. For \object{KIC~4247791}, the periods of $4.10$ and $4.05$\,days are clearly off the 3:2 ratio, although their relative difference is again $\sim 10^{-2}$. The nearly resonant periods in CzeV343 may very well be a coincidence, especially if there is no physical connection between the binaries. There is no theoretical basis for such a behavior neither, but to our knowledge, there has not been any study of KCTF-driven evolution of a quadruple system. However, with the current interest in astrophysical Kozai cycles \citep[e.g.][]{blaes02,fabrycky07,karas07,perets09,thompson11,dong12b} expanding the studies to binary pairs may be a logical extension.

\begin{table}
\caption{Summary of observations.}
\label{tab:nights}
\centering
\renewcommand{\arraystretch}{1.2}
\begin{tabular}{cccccc}
\hline\hline
$t_{\rm start}$\tablefootmark{a} & $\Delta t$\tablefootmark{b} & Filter\tablefootmark{c} & $N_j$ & $\Delta T_{0,A}$\tablefootmark{d} & $\Delta T_{0,B}$\tablefootmark{d}\\
\hline
5957.352 &  2.86 & $C$ & 40 &                       &     $-0.0^{+ 1.6}_{-1.9}$\tablefootmark{f} \\
5958.328 &  4.39 & $C$ & 61 &     $-1.3^{+ 1.4}_{-1.3}$\tablefootmark{e} &                       \\
5959.293 &  4.68 & $C$ & 63 &                       &     $-3.4^{+ 3.9}_{-3.4}$\tablefootmark{e} \\
5968.273 &  5.26 & $C$ & 71 &                       & \phs$ 1.4^{+ 1.6}_{-1.5}$\tablefootmark{e} \\
5970.289 &  4.58 & $C$ & 57 &     $-1.8^{+ 0.8}_{-0.7}$\tablefootmark{e} & \phs$ 0.4^{+ 1.4}_{-1.1}$\tablefootmark{f} \\
5984.320 &  5.16 & $C$ & 74 & \phs$ 7.8^{+ 1.4}_{-1.3}$\tablefootmark{f} & \phs$ 0.2^{+ 1.2}_{-1.2}$\tablefootmark{e} \\
5960.305 &  3.55 & $C$ & 34 &                       &                       \\
5990.254 &  5.21 & $C$ & 76 & \phs$ 4.5^{+ 2.0}_{-1.8}$\tablefootmark{f} &                       \\
5992.254 &  4.82 & $C$ & 85 &                       &                       \\
5993.305 &  3.79 & $C$ & 67 &     $-2.4^{+ 1.0}_{-1.1}$\tablefootmark{e} &     $-2.1^{+ 1.5}_{-1.7}$\tablefootmark{e} \\
6008.285 &  3.22 & $C$ & 60 &                       &                       \\
6012.297 &  2.35 & $C$ & 68 &                       &     $-0.3^{+ 0.9}_{-0.9}$\tablefootmark{f} \\
6027.277 &  2.59 & $I_{\rm C}$ & 20 &     $-0.3^{+ 2.5}_{-1.5}$\tablefootmark{e} &                       \\
6027.273 &  2.59 & $V$ & 20 &     $-2.4^{+ 2.2}_{-1.6}$\tablefootmark{e} &                       \\
6035.293 &  2.45 & $I_{\rm C}$ & 17 &                       &     $-7.0^{+ 5.6}_{-5.8}$\tablefootmark{e} \\
6035.289 &  2.59 & $V$ & 20 &                       &     $-3.8^{+ 3.8}_{-3.2}$\tablefootmark{e} \\
\hline
\end{tabular}
\tablefoot{Each line corresponds to a single dataset.
\tablefoottext{a}{Beginning of observations in BJD$_{\rm TDB} - 2\,450\,000$.}
\tablefoottext{b}{Duration of observations in hours.}
\tablefoottext{c}{Passband of the dataset. $C$ stands for unfiltered observations while $V$ and $I_{\rm C}$ are standard Johnson-Cousins filters.}
\tablefoottext{d}{The fitted minus mean value of $T_0$ in the units of $10^{-3}$\,days. The uncertainties are $90\%$ confidence intervals. }
\tablefoottext{e}{Primary minimum.}
\tablefoottext{f}{Secondary minimum.}
}
\end{table}

\begin{table}
\caption{Parameters of CzeV343.}
\label{tab:params}
\centering
\renewcommand{\arraystretch}{1.2}
\begin{tabular}{ccc}
\hline\hline
Parameter & System A & System B \\
\hline
$P$ [d] & $1.209373^{+0.000019}_{-0.000017}$ & $0.806931^{+0.000016}_{-0.000019}$ \\
$T_0$\tablefootmark{a}   & $5958.36058^{+0.00042}_{-0.00045}$ & $5968.33977^{+0.00047}_{-0.00045}$ \\
$i$ [deg] &  $90.0^{+0.0}_{-1.1}$ &  $67.11^{+0.60}_{-0.29}$ \\\
$e\cos\omega$ & $0.0147^{+0.0010}_{-0.0009}$ & $\equiv 0$\tablefootmark{b} \\
$e\sin\omega$ & $0.178^{+0.014}_{-0.013}$ & $\equiv0$\tablefootmark{b} \\
$\sigma$ &  $0.590^{+0.013}_{-0.010}$ & $0.989^{+0.047}_{-0.027}$\\
$r_1+r_2$ & $0.4533^{+0.0065}_{-0.0046}$ & $0.581^{+0.005}_{-0.012}$\\
$r_2/r_1$ & $1.267^{+0.015}_{-0.017}$ & $1.05^{+0.21}_{-0.05}$\\
\hline
\end{tabular}
\tablefoot{
The uncertainties are $90\%$ confidence intervals. 
\tablefoottext{a}{BJD$_{\rm TDB} - 2\,450\,000$.}
\tablefoottext{b}{Fixed during fitting.}
}
\end{table}

\begin{acknowledgements}
We thank Todd Thompson, Chris Kochanek and Kris Stanek for advice, encouragement and detailed reading of the manuscript. We are grateful to Ben Shappee, Scott Gaudi, Ji\v{r}\'{i} Li\v{s}ka, and John Southworth for comments and discussions. This research has made use of SDSS. Funding for the SDSS and SDSS-II has been provided by the Alfred P. Sloan Foundation, the Participating Institutions, the National Science Foundation, the U.S. Department of Energy, the National Aeronautics and Space Administration, the Japanese Monbukagakusho, the Max Planck Society, and the Higher Education Funding Council for England. The SDSS Web Site is http://www.sdss.org/.
\end{acknowledgements}

\bibliographystyle{aa}
\end{document}